\documentclass[twocolumn,aps,prc,showpacs,floatfix]{revtex4}
\usepackage{graphicx}
\usepackage{amsmath}
\usepackage{dcolumn}
\usepackage{bm}

\begin{document}

\preprint{}
\title{Effects of the high-momentum tail of nucleon momentum distribution on the isospin-sensitive observables}
\author{Fang Zhang$^{1}$}\email{zhangfang@lzu.edu.cn}
\author{Gao-Chan Yong$^{2}$\footnote{Corresponding author}}\email{yonggaochan@impcas.ac.cn}
\affiliation{%
$^1${School of Nuclear Science and Technology, Lanzhou
University, Lanzhou 730000, People's Republic of China}\\
$^2${Institute of Modern Physics, Chinese Academy of Sciences, Lanzhou
730000, China}
}%


\begin{abstract}

Based on the Isospin-dependent transport model Boltzmann-Uehling-Uhlenbeck (IBUU),
effects of the high momentum tail (HMT) of nucleon momentum distribution in colliding nuclei on some isospin-sensitive observables are studied in the semi-central $^{197}\rm {Au}+^{197}\rm {Au}$ reactions at incident beam energy of 400 MeV/nucleon. It is found that the nucleon transverse flow, the difference of neutron and proton transverse flows, the nucleon elliptic flow and free neutron to proton ratio are all less sensitive to the HMT,
while the isospin-sensitive nucleon elliptic flow difference is clearly affected by the HMT.
Except at very high kinetic energies, the kinetic energy distributions of $\pi^{-}$, $\pi^{+}$ and charged pion ratio $\pi^{-}/\pi^{+}$ are all sensitive to the HMT.

\end{abstract}

 \maketitle

\section{Introduction}

The results of recent high-momentum transfer reactions demonstrate that about $20\%$ nucleons form strongly correlated nucleon pairs with large relative momentum and small center-of-mass momentum (relative to the Fermi momentum) \cite{E06, R07,sci08}. Such phenomenon is explained by the short range nucleon-nucleon interactions \cite{M05,Schiavilla07,Rios14,XC13}. And such short range nucleon-nucleon interaction is strong in neutron-proton pairs but weak in neutron-neutron
or proton-proton pairs. Thus in isospin asymmetric nuclei, protons have a larger probability than neutrons to have momenta higher
than the Fermi momentum \cite{sci14, Sargsian12}.

Nowadays the equation of state (EoS)
of isospin symmetric nuclear matter is relatively well
determined but the EoS of isospin
asymmetric nuclear matter, especially the density dependence
of the nuclear symmetry energy is still uncertain \cite{Bar05,LCK08,pawl2002,Guo14}. Besides being of great importance in nuclear physics, the symmetry energy
also plays crucial roles in many astrophysical processes. Thus the crucial task is to find experimental observables that are sensitive to the symmetry energy. A number of potential observables have been identified in heavy-ion collisions induced by neutron-rich
nuclei, such as the free neutron/proton ratio \cite{liba97}, the $\pi^-/\pi^+$ ratio \cite{yong2006}, the t/$^{3}$He \cite{yongflow09}, the nucleon transverse flow \cite{yonggc06}, the eta production \cite{eta13}, etc. More related references can be found in Ref.~\cite{Bar05,LCK08,Guo13,Guo14}.

Because both isospin-sensitive nucleon transverse and elliptic flows depend on nucleon momentum distribution, they are expected to be sensitive to the HMT. In addition, in heavy-ion collisions the primordial
$\pi^{-}/\pi^{+}$ ratio is approximately equal to
$(N/Z)^{2}$ at certain momentum space and the $\pi^{-}$'s are mostly produced
from neutron-neutron collisions while $\pi^{+}$'s are mostly produced
from proton-proton collisions \cite{LiBA02,stock86}. Thus the kinetic energy distribution of the $\pi^{-}/\pi^{+}$ ratio is expected to be also sensitive to the HMT.

Since the HMT of single-nucleon momentum distribution in nucleus was confirmed experimentally at Jefferson Laboratory, it should be
considered in the nuclear transport simulations. Unfortunately, most isospin-dependent transport models rarely took the effects of the HMT into account \cite{D07, M09, W09, Xiao09, M13}.
Recently, the effects of the HMT have been added into the transport model by modifying the symmetry potential in nuclear mean-field potential \cite{henprc15,balig15}.
It is also necessary to see the HMT effects of nucleon initial momentum distribution of nuclei on some frequently studied isospin-sensitive observables in heavy-ion collisions, such as nucleon collective flow, free neutron to proton ratio $n/p$ as well as the $\pi^{-}/\pi^{+}$ ratio. We make these studies in $^{197}\rm {Au}+^{197}\rm {Au}$ collisions at a beam energy of 400 MeV/nucleon with an impact parameter b = 7.2 fm.
The reasons of selecting b = 7.2 fm are that there are FOPI-LAND data for Au+Au
at 400 MeV/nucleon with b = 7.2 fm. And also with larger impact parameters, one can get clear nucleon flow.

Our paper is organized as follows. The next section gives a brief description of the applied IBUU model. The following is the detailed studies of the HMT effects on
the collective flow and the free neutron to proton ratio $n/p$ and the $\pi^{-}/\pi^{+}$ ratio. The conclusion is given in the last section.

\section{THE IBUU TRANSPORT MODEL}

\begin{figure}[t]
\centering
\includegraphics[width=0.5\textwidth]{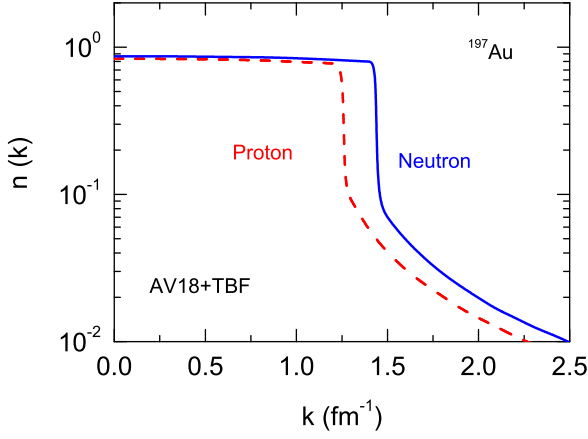}
\caption{ Momentum distributions of neutron and proton in nucleus $^{197}$\rm {Au}. Taken from Ref.~\cite{yin13}.} \label{npdis}
\end{figure}
To study the effects of the HMT of nucleon momentum distribution on isospin-sensitive observables,
we use the Isospin-dependent
Boltzmann-Uehling-Uhlenbeck (IBUU) transport model \cite{ib96,ib97,ib99,LiBA02,liba2003,yong20151,yong20152}.
In this IBUU model, nucleon coordinates
are given in nucleus with radius R = 1.2 A$^{1/3}$, where A is the mass number of nucleus.
Fig.~\ref{npdis} shows nucleon initial momentum distribution with high-momentum tail in $^{197}$\rm {Au}, which is given by the extended brueckner-hartree-fock (BHF)
approach by adopting the AV 18 two-body interaction plus a microscopic three-body-force (TBF) \cite{yin13}. The instability of initial
colliding nuclei including the HMT has negligible effects on our studies here.
As comparison, we also give nucleon momentum distribution
\begin{eqnarray}
n(k)=\left\{%
  \begin{array}{ll}
    1, & \hbox{$k \leq k_{F}$;} \\
    0, & \hbox{$k > k_{F}$.} \\
\end{array}%
\right.
\end{eqnarray}

In this model, we use the isospin- and momentum-dependent mean-field
single nucleon potential \cite{Das03,xu14,yong20151,yong20152}
\begin{eqnarray}
U(\rho,\delta,\vec{p},\tau)&=&A_u(x)\frac{\rho_{\tau'}}{\rho_0}+A_l(x)\frac{\rho_{\tau}}{\rho_0}\nonumber\\
& &+B(\frac{\rho}{\rho_0})^{\sigma}(1-x\delta^2)-8x\tau\frac{B}{\sigma+1}\frac{\rho^{\sigma-1}}{\rho_0^\sigma}\delta\rho_{\tau'}\nonumber\\
& &+\frac{2C_{\tau,\tau}}{\rho_0}\int
d^3\,\vec{p^{'}}\frac{f_\tau(\vec{r},\vec{p^{'}})}{1+(\vec{p}-\vec{p^{'}})^2/\Lambda^2}\nonumber\\
& &+\frac{2C_{\tau,\tau'}}{\rho_0}\int
d^3\,\vec{p^{'}}\frac{f_{\tau'}(\vec{r},\vec{p^{'}})}{1+(\vec{p}-\vec{p^{'}})^2/\Lambda^2},
\label{buupotential}
\end{eqnarray}
where $\tau, \tau'=1/2(-1/2)$ for neutrons (protons), $\rho_0$ is nuclear saturation density.
$\delta=(\rho_n-\rho_p)/(\rho_n+\rho_p)$ is the isospin asymmetry,
and $\rho_n$, $\rho_p$ denote neutron and proton densities,
respectively. The parameter values $A_u(x)$ = 33.037 - 125.34$x$
MeV, $A_l(x)$ = -166.963 + 125.34$x$ MeV, B = 141.96 MeV,
$C_{\tau,\tau}$ = 18.177 MeV, $C_{\tau,\tau'}$ = -178.365 MeV, $\sigma =
1.265$, and $\Lambda = 630.24$ MeV/c are obtained by fitting
empirical constraints of the saturation density, the binding energy, the
incompressibility, the isoscalar effective mass, the single-particle potential, the symmetry
energy value and the symmetry potential
at infinitely large nucleon momentum
at saturation density. $f_{\tau}(\vec{r},\vec{p})$ is
the phase-space distribution function at coordinate $\vec{r}$ and
momentum $\vec{p}$ and solved by using the test-particle method
numerically. The symmetry energy's stiffness parameter $x$
in the above single nucleon potential is used to mimic
different forms of the symmetry energy.
Since we do not intend to study the effect of nuclear symmetry energy,
here we just let $x$ = 0. Note here that, besides considering the HMT in momentum initialization,  the effect of the HMT on the single particle
potential is also considered consistently \cite{yong20152}.
For $\Delta$ baryon potential,
it is divided into nucleon
potential by \cite{LiBA02,GYZ15}
\begin{eqnarray}
U_{\Delta^-}&=& U(\rho,\delta,\vec{p},\tau = \frac{1}{2}),\\
U_{\Delta^0}&=& \frac{2}{3}U(\rho,\delta,\vec{p},\tau = \frac{1}{2})+\frac{1}{3}U(\rho,\delta,\vec{p},\tau = -\frac{1}{2}),\\
U_{\Delta^+}&=& \frac{1}{3}U(\rho,\delta,\vec{p},\tau = \frac{1}{2})+\frac{2}{3}U(\rho,\delta,\vec{p},\tau = -\frac{1}{2}),\\
U_{\Delta^{++}}&=& U(\rho,\delta,\vec{p},\tau = -\frac{1}{2}).
\end{eqnarray}
We use the reduced baryon-baryon ($BB$) scattering cross section in medium by a factor of
\begin{eqnarray}
R_{\rm {medium}}(\rho,\delta,\vec{p})&\equiv& \sigma
_{BB_{\rm {elastic, inelastic}}}^{\rm {medium}}/\sigma
_{BB_{\rm {elastic, inelastic}}}^{\rm {free}}\nonumber\\
&=&(\mu _{BB}^{\ast }/\mu _{BB})^{2},
\end{eqnarray}
where $\mu _{BB}$ and $\mu _{BB}^{\ast }$ are the reduced masses
of the colliding baryon-pair in free space and medium,
respectively. Similar with the definition in Ref.~\cite{liba2003},
the effective mass of baryon in isospin asymmetric nuclear matter
is expressed by
\begin{equation}
\frac{m_{B}^{\ast }}{m_{B}}=\left\{ 1+\frac{m_{B}}{p}\frac{%
dU_{B}}{dp}\right\},
\end{equation}
where $dU_{B}$ denotes nucleon or $\Delta$ potentials.

\section{Results and discussion}

\begin{figure}[th]
\begin{center}
\includegraphics[width=0.5\textwidth]{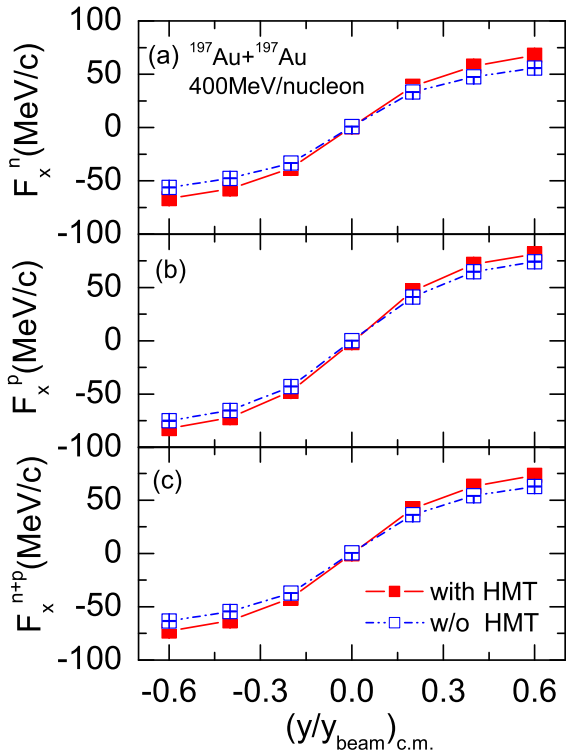}
\end{center}
\caption {(Color online) Rapidity distribution of nucleon transverse flow (upper: neutron, middle: proton, bottom: nucleon) in the semi-central reaction $^{197}\rm {Au}+^{197}\rm {Au}$ at incident beam energy of $400$ MeV/nucleon, with and without the HMT, respectively.}
\label{pxA}
\end{figure}
Nucleon collective transverse flow is the moving particles deflected away from the beam axis in the reaction plane, which reads \cite{ib99,flow85,flow86,flow88,flow93,gao12}
\begin{equation}\label{}
      F(y)=\frac{1}{N(y)}\sum_{i=1}^{N(y)}p_{i}^{x}(y).
\end{equation}
In the above, $p_{x}$ is the momentum of an emitting nucleon in $x$ direction at
rapidity $y$, $N(y)$ is the number of free nucleons at rapidity $y$.
Here, free nucleons are identified when their local densities are less than $\rho_{0}/8$.
Shown in Fig.\ \ref{pxA} is the effects of the HMT on the nucleon transverse flow as a function of reduced rapidity. From Fig.\ \ref{pxA}, it is seen that the effects of the HMT on the transverse flow are in fact not evident. This is understandable since the shape of
high-momentum tail roughly exhibits
a $C/k^{4}$ form \cite{sci14,henprc15}, the number of nucleons with momenta larger than the Fermi momentum
is just a small proportion of the total number of nucleons in a nucleus. Nevertheless, the strength of the nucleon transverse flow becomes somewhat larger with the HMT than that without the HMT.

In the high-momentum tail of nucleon momentum distribution,
since nucleonic component is strongly isospin-dependent, i.e., the
number of n-p pairs is about 18 times that of the p-p or n-n
pairs \cite{sci08}, one expects larger effects of the HMT on the isospin-sensitive collective flow.

\begin{figure}[th]
\begin{center}
\includegraphics[width=0.5\textwidth]{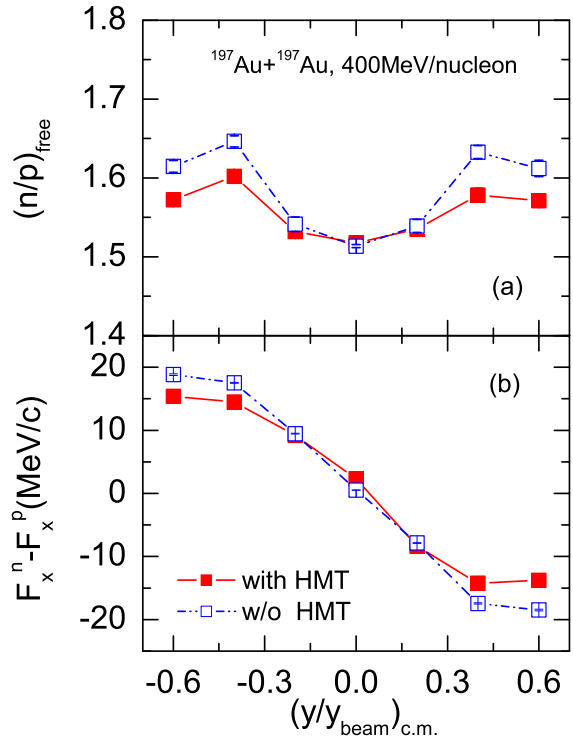}
\end{center}
\caption{(Color online) Rapidity distribution of free neutron to proton ratio
(upper panel), the difference of neutron and proton transverse flows (lower panel)
in the semi-central $^{197}\rm {Au}+^{197}\rm {Au}$ reaction at incident beam energy of
$400$ MeV/nucleon with and without the HMT.}
\label{pxnp}
\end{figure}
Fig.\ \ref{pxnp} shows the rapidity distribution of the
free neutron to proton ratio $n/p$  and the difference
of neutron and proton transverse flows. Since n-p pairs dominate
in the HMT, from Fig.\ \ref{pxnp}, one sees lower value
of the $n/p$ ratio at larger rapidities with the HMT. And also with the HMT, relatively small absolute value of the difference of neutron and proton flows $(F_{x}^{n}(y)-F_{x}^{p}(y))$ at larger rapidities is seen in the lower panel. From Fig. 2 - 3, one can see that the effects of the HMT on the free neutron to proton ratio $n/p$ and the nucleon transverse flow are clear but not large. This is really good news for some physical goals such as studying nuclear symmetry energy by these observables in heavy-ion collisions. Nevertheless, one has to see if other isospin-sensitive observables are sensitive to the HMT.

\begin{figure}[th]
\begin{center}
\includegraphics[width=0.5\textwidth]{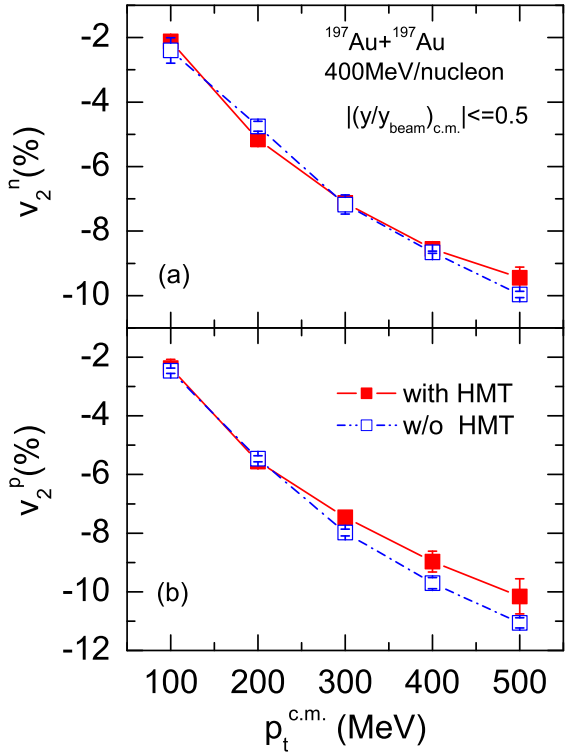}
\end{center}
\caption{(Color online) Transverse momentum distribution of neutron (upper panel) and proton (lower panel) elliptic flows in the semi-central $^{197}\rm {Au}+^{197}\rm {Au}$ collisions
at a beam energy of 400 MeV/nucleon with and without the HMT, respectively.}
\label{v2ptp}
\end{figure}
We now turn to study the effects of the HMT on the differential elliptic
flow of unbound nucleons. The elliptic flow corresponds to the second
Fourier coefficient in the transverse-momentum distribution and can be expressed as \cite{ib99,gao12,ditorof,ditorof2}
\begin{equation}\label{}
  v_{2}=\langle \frac{p_{x}^{2}-p_{y}^{2}}{p_{x}^{2}+p_{y}^{2}}\rangle.
\end{equation}
Where $p_{x}$ is the nucleon transverse momentum along $x$ axis in the reaction plane, $p_{y}$ is the
nucleon transverse momentum along $y$ axis perpendicular to the reaction plane.
A negative value of $v_{2}$ describes the dominant out-of-plane particle emission
while a positive value $v_{2}$ indicates the leading in-plane particle emission. Shown in Fig.\ \ref{v2ptp} is the transverse momentum distribution of nucleon elliptic flow at mid-rapidity with or without the HMT.
It is seen that the values of both neutron and proton elliptic flows are negative, which indicating the leading out-of-plane nucleon emission at such incident beam energy. The strength of proton elliptic flow is larger than that of neutron, which indicating a strong Coulomb repulsion to protons. And one can also see that the effect of the HMT on the proton elliptic flow is larger than that of neutron. This is because in neutron-rich reaction system, protons have great probabilities than neutrons to be affected by the short-range correlations \cite{sci14}. Anyway, the effects of the HMT on nucleon elliptic flow are not large in the low transverse momentum region but somewhat larger in the high transverse momentum region.

\begin{figure}[th]
\begin{center}
\includegraphics[width=0.5\textwidth]{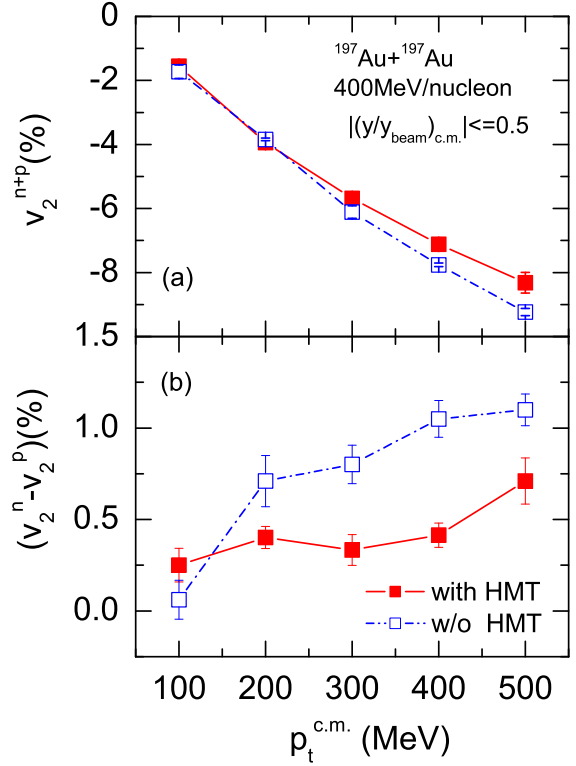}
\end{center}
\caption{(Color online) Total nucleon elliptic flow (upper panel) and the
difference of neutron and proton elliptic flow (lower panel) in the semi-central reaction
of $^{197}\rm {Au}+^{197}\rm {Au}$ at a beam energy of 400 MeV/nucleon with
and without the HMT, respectively.}
\label{v2ptnp}
\end{figure}
\begin{figure}[th]
\begin{center}
\includegraphics[width=0.5\textwidth]{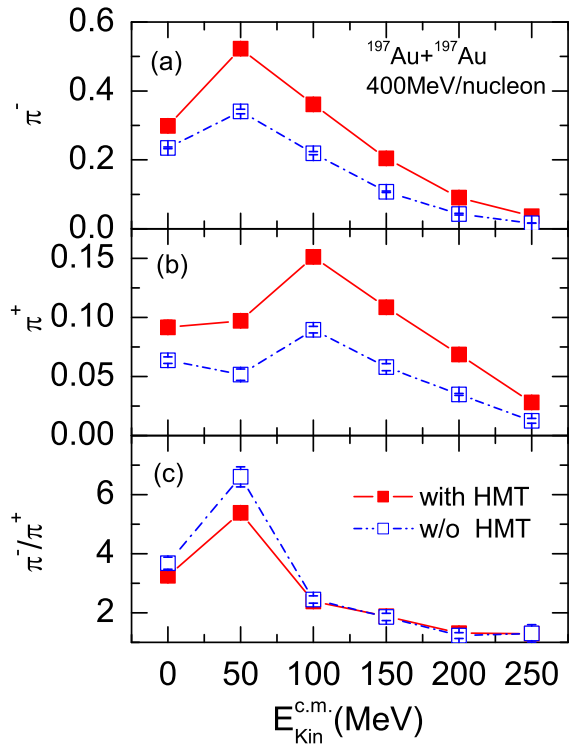}
\end{center}
\caption{(Color online) Kinetic energy distributions of $\pi^{-}$ (upper panel),
$\pi^{+}$ (middle panel) as well as $\pi^{-}/\pi^{+}$ (bottom panel) ratio with and without the HMT in the reaction of $^{197}\rm {Au}+^{197}\rm {Au}$ at incident beam energy of 400 MeV/nucleon.}
\label{Rpion}
\end{figure}
Since the effects of the HMT on nucleon elliptic flow are generally small, we expect the
effects of the HMT on the
difference of neutron and proton elliptic flows ($v_{2}^{n}-v_{2}^{p}$) to be larger. For comparison, the total nucleon elliptic
flow $v_{2}^{n+p}$ is also presented in the upper panel of Fig.\ \ref{v2ptnp}.
Shown in the lower panel of Fig.\ \ref{v2ptnp}, the effects of the HMT on the difference of
neutron and proton elliptic flows ($v_{2}^{n}-v_{2}^{p}$) is larger than that of the total nucleon elliptic flow. Because the proton elliptic flow is negative and stronger than the neutron elliptic flow, the difference of neutron and proton elliptic flows ($v_{2}^{n}-v_{2}^{p}$) is positive and becomes smaller due to neutron-proton correlations in the HMT. Comparing Fig.\ \ref{v2ptnp} with Fig.\ \ref{pxnp}, it is found that
the isospin-sensitive elliptic flow observable ($v_{2}^{n}-v_{2}^{p}$) is sensitive to the HMT.

Finally, we examine in Fig.\ \ref{Rpion} the kinetic energy distributions of $\pi^{-}$, $\pi^{+}$ as well as $\pi^{-}/\pi^{+}$ ratio with and without the HMT in the same semi-central reaction
$^{197}\rm {Au}+^{197}\rm {Au}$ at a beam energy of 400 MeV/nucleon (while simulating central collisions
of $^{197}\rm {Au}+^{197}\rm {Au}$ at the same beam energy, we obtained the same physical results).
It is seen that kinetic energy distributions of both $\pi^{-}$ and $\pi^{+}$ are very sensitive to the HMT. The ratio of
$\pi^{-}/\pi^{+}$ is also very sensitive to the HMT except in the high kinetic energy region.
The neutron-proton short-range corrections increase kinetic energies of a certain proportion of  neutrons and protons, thus more pion mesons are produced. This is the reason why the value of the kinetic energy distribution of pion meson is higher with the HMT than that without the HMT. Because in isospin asymmetric reaction system, protons have a larger probability than neutrons to have larger momenta \cite{sci14, Sargsian12} and proton-proton collision mainly produces $\pi^{+}$, one sees lower value of the $\pi^{-}/\pi^{+}$ ratio with the HMT than that without the HMT. The effects of the HMT on the $\pi^{-}/\pi^{+}$ ratio become insensitive in the high kinetic region. However, in the high-energy tail of $\pi$ spectra, the $\pi^{-}/\pi^{+}$ ratio could be even more sensitive to the symmetry energy at high densities \cite{gaoy13,junh14}. This is surely interesting to those who use the observable $\pi^{-}/\pi^{+}$ ratio to probe the symmetry energy \cite{LiBA02,yong06,Gai04,LiQF05b,junh14,ko15,xie13} without including the HMT in their transport models.

\section{Summary}

In the framework of the isospin-dependent transport model, we studied the effects of the high momentum tail of nucleon momentum distribution in colliding nuclei on some isospin-sensitive observables. We find that in the same semi-central reaction
$^{197}\rm {Au}+^{197}\rm {Au}$ at a beam energy of 400 MeV/nucleon, the isoscalar nucleon flows, including transverse and elliptic flows are less affected by the high momentum tail of nucleon momentum distribution in colliding nuclei. The isovector nucleon elliptic flow is sensitive to the high-momentum tail of nucleon momentum. Except the energetic $\pi^{-}/\pi^{+}$ ratio, the kinetic energy distributions of $\pi^{-}$, $\pi^{+}$ and charged pion ratio $\pi^{-}/\pi^{+}$ are all sensitive to the high momentum tail of nucleon momentum distribution.

Because the nucleon-nucleon correlation is hard to embed into the transport model, to probe the symmetry energy using isospin-sensitive observables, the HMT-insensitive observable such as the difference of neutron and proton transverse flows \cite{yonggc06,gao12,ditorof,ditorof2} and $\pi^{-}/\pi^{+}$ ratio at higher energies \cite{gaoy13,junh14} are recommendable.
And also because the HMT itself is isospin-dependent, it is necessary to further study the competitive relation of the HMT and the symmetry energy on frequently used isospin-sensitive observables.

\section*{Acknowledgments}

The work is supported partly by the Fundamental Research Funds for the Central Universities (lzujbky-2014-170) and the National Natural Science Foundation of China (Grant Nos. 11375239, 11435014).

\end{document}